\documentclass[useAMS,usenatbib]{mn2e}
\usepackage{graphicx}
\usepackage{natbib}  
\usepackage{epstopdf}
\newcommand\aj{{AJ}}%
\newcommand\araa{{ARA\&A}}%
\newcommand\apj{{ApJ}}%
\newcommand\apjl{{ApJ}}%
\newcommand\apjs{{ApJS}}%
\newcommand\aap{{A\&A}}%
\newcommand\aaps{{A\&AS}}%
\newcommand\mnras{{MNRAS}}%
\def\simgt{\lower.5ex\hbox{$\; \buildrel > \over \sim \;$}}
\def\simlt{\lower.5ex\hbox{$\; \buildrel < \over \sim \;$}}

\newcommand\teff{T$_{\rm eff}$}

\newcommand{\msun}{\ensuremath{\, {M}_\odot}}
\newcommand{\Msun}{\ensuremath{\, {M}_\odot}}
\newcommand{\ocen}{$\omega$~Cen}

\title[]{The C+N+O abundances and the splitting of the subgiant branch in the
Globular Cluster NGC~1851}
\author[P. Ventura, V. Caloi, F. D'Antona, J. Ferguson, A. Milone \& G.Piotto]{P. Ventura$^{1}$, 
V.Caloi$^{2}$, F. D'Antona$^{1,2}$, J. Ferguson$^{3}$, A. Milone$^4$\ and G.P. Piotto$^4$
\thanks{E-mail: ventura@oa-roma.inaf.it (PV); dantona@oa-roma.inaf.it (FD),
vittoria.caloi@iasf-roma.inaf.it (VC); antonino.milone@ unipd.it (AM), 
giampaolo.piotto@ unipd.it (GP);
jason.ferguson@wichita.edu (JF)}
\\
$^{1}$ INAF, Osservatorio Astronomico di Roma, Via Frascati 33, 
00040 Monteporzio Catone (Roma), Italy.\\
$^{2}$ INAF, IASF--Roma, via Fosso del Cavaliere 100, I-00133 Roma, Italy\\
$^{3}$ Department of Physics, Wichita State University, Wichita KS 67260-0032, USA\\
$^{4}$ Dipartimento di Astronomia, Universit\`a di Padova, 
Vicolo dellOsservatorio 3, Padova, I-35122, Italy
}
\begin{document}

\date{Accepted . Received ; in original form }

\pagerange{\pageref{firstpage}--\pageref{lastpage}} \pubyear{2006}

\maketitle

\label{firstpage}

\begin{abstract}
Among the newly discovered features of multiple stellar populations in Globular Clusters, 
the cluster NGC~1851 harbours a double subgiant branch, that can be explained in terms of
two stellar generations, only slightly differing in age, the younger one having an
increased total C+N+O abundance. Thanks to this difference in the chemistry, 
a fit can be made to the subgiant branches, roughly consistent with the C+N+O abundance 
variations already discovered two decades ago, and confirmed by recent spectroscopic data. 
We compute theoretical isochrones for the main sequence turnoff, by adopting four chemical 
mixtures for the opacities and nuclear reaction rates. 
The standard mixture has Z=10$^{-3}$ and [$\alpha$/Fe]=0.4, the others have
C+N+O respectively equal to 2, 3 and 5 times the standard mixture, according to the element
abundance distribution described in the text.
We compare tracks and isochrones, and show how the results depend on the total CNO
abundance. We notice that different initial 
CNO abundances between two clusters, otherwise similar in metallicity and age,
may lead to differences in the turnoff morphology that can be easily attributed
to an age difference.
We simulate the main sequence and subgiant branch data for NGC~1851 and show that 
an increase of C+N+O by a factor $\sim$3 best reproduces the shift between the subgiant 
branches. According to spectroscopic data by Yong et al., the C+N+O abundance in this 
cluster appears correlated with the abundance of s-process elements, Na and Al,
and this makes massive AGBs the best progenitors of the C+N+O enriched population. 
We compare the main sequence width in the color m$_{F336W}$-m$_{F814W}$ with models,
and find that the maximum helium abundance compatible with the data is Y$\simeq$0.29.
We consider the result in the framework of the formation of the second 
stellar generation in globular clusters,
for the bulk of which we estimate a helium abundance of Y$\simlt 0.26$. The precise value
depends on which are the AGB masses from which the C+N+O enriched matter originates, 
and on the amount of dilution with the pristine gas.
\end{abstract}

\begin{keywords}
globular clusters:general; globular clusters:individual: NGC~1851; stars:abundances
\end{keywords}

\section{Introduction}
\label{sec:intro}

\begin{figure}
\includegraphics[width=8cm]{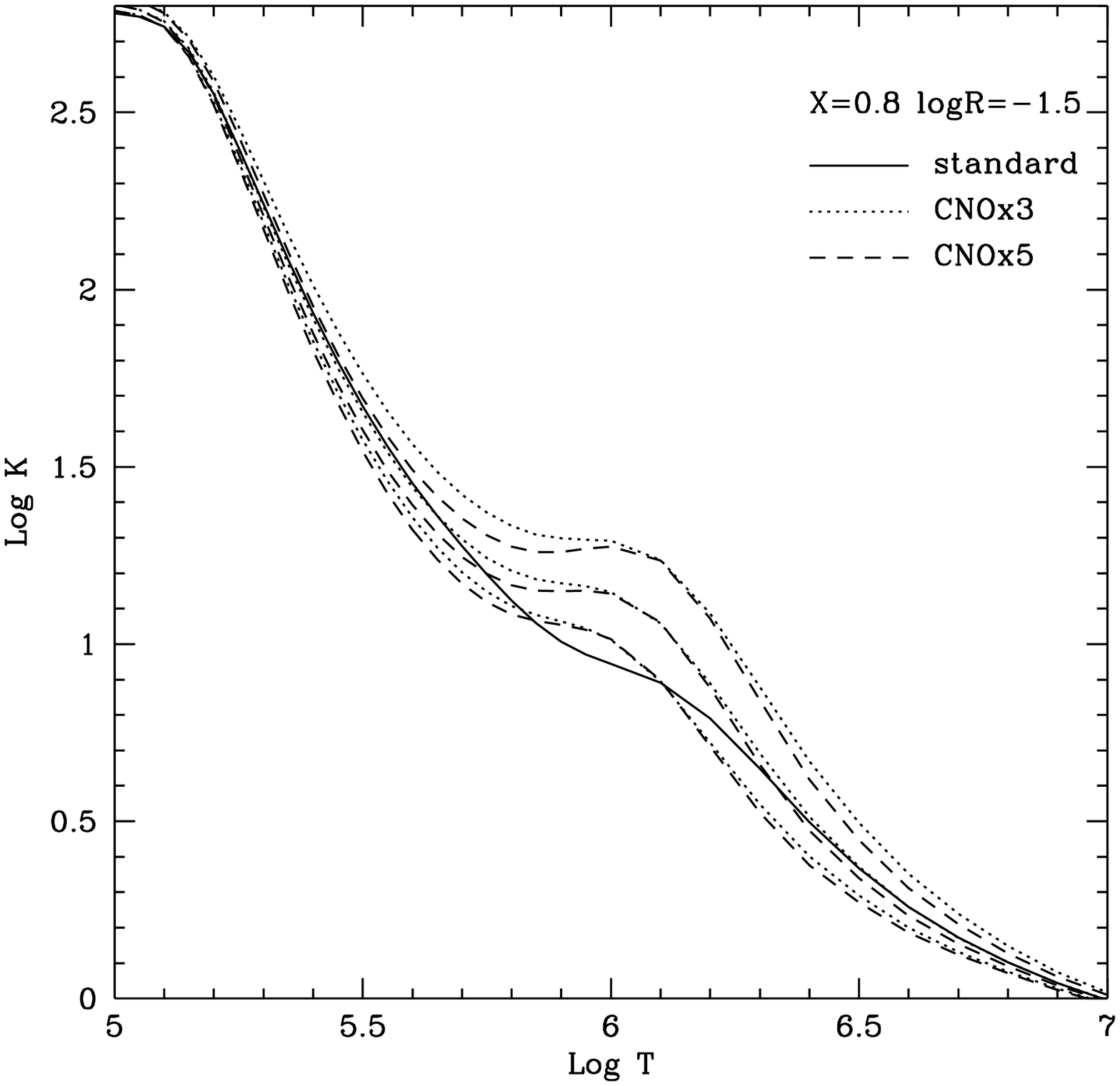}
\includegraphics[width=8cm]{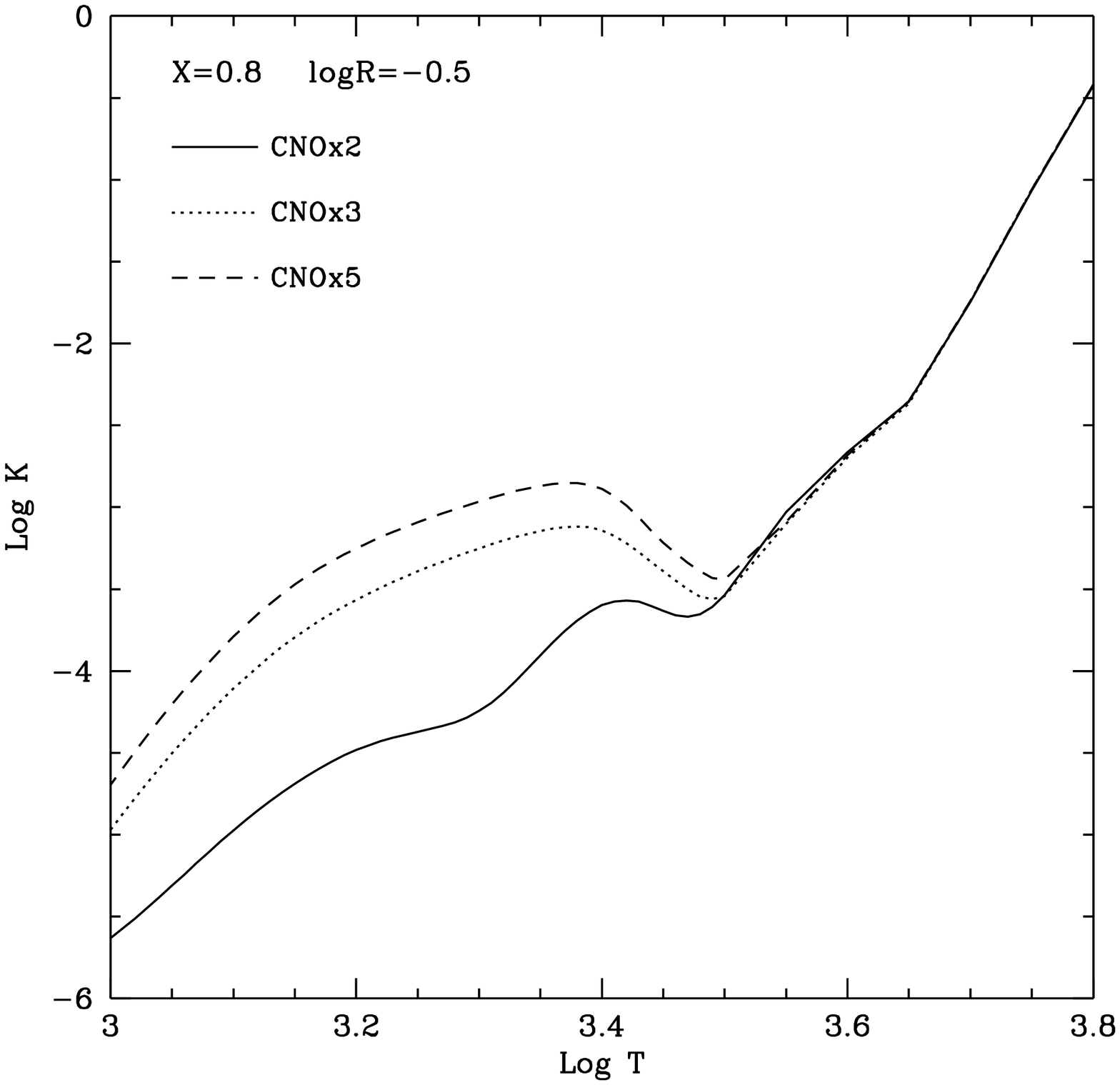}
\caption{Variation with temperature of the opacity of the 
mixtures used in the present investigation. 
At the top of the figures, the value of R=density/temperature$^3$ and of the
hydrogen mass fraction X are labelled. Upper panel: opacity
values for the standard (solid track), $\alpha-$ enhanced case, and for 
the mixtures with CNO abundances increased by
a factor 3 (dotted) and 5 (dashed), corresponding to a total
metallicity Z=0.00235 and Z=0.00350. For the two latter cases
we plot the values for three total metallicities Z=1, 2, and 4$\times 10^{-3}$ 
(from left to right).
Lower panel: opacities in the low-temperature regime, for the mixtures
with CNO enhancements of 2 (solid), 3 (dotted), and 5 (dashed).
}
      \label{opacity}%
\end{figure}

The observations of Globular Cluster (GC) stars still need to be interpreted in
a fully consistent frame.  Nevertheless, a general consensus is emerging on
the fact that most GCs can not be considered any longer simple stellar
populations, and that ``self--enrichment" is a common feature among
them. The first suspicions originated from the well known ``chemical anomalies'',
already noted in the seventies (such as the Na--O and Mg--Al
anticorrelations). Recently observed to be present at the turnoff (TO) and
among the subgiants \citep[e.g.][]{gratton2001,briley2002, briley2004}, they
must be attributed to some process of self--enrichment occurring at the
first stages of the cluster life. 
A decisive feature indicating the presence of more than one population in GCs 
has been the observation of multiple, well separated sequences. Multiple
main sequences have been found in \ocen\ \citep{bedin2004, piotto2005}
and NGC 2808 \citep{dantona2005, piotto2007}. Besides, multiple subgiant branches have 
been observed in NGC~1851 \citep{milone2008} 
and other clusters, among which NGC~6388 \citep{piotto2009}.
The former phenomenon can only be interpreted in terms of populations with different 
helium content \citep{dantona2002,norris2004}. The latter may be discussed in terms
of a difference in age \citep{milone2008}, or in total CNO abundance \citep{cassisi2008},
and is the subject of the present work. In both cases, we are left with the difficulty of finding a 
consistent origin for these populations. We can speculate that there 
was a first epoch of star
formation, that gave origin to the ``normal" (first generation, hereinafter FG)
stars, with CNO and other abundances similar to Population II field stars of
the same metallicity. Afterwards, there must have been some other epoch of
star formation (second generation, hereinafter SG), including material heavily
processed through the hot CNO cycle in the progenitors, belonging to the FG, 
but not enriched in the heavy elements expected in supernova ejecta.
This material either comes entirely from the
stars belonging to the first stellar generation, or it is a mixture of processed
gas and pristine matter of the initial star forming cloud. 

The SG in most clusters is a high fraction of the total number of stars \citep{carretta2008,
dantonacaloi2008}. In order to have enough CNO processed material available, it is necessary
that ``self--enrichment" is a result of pollution from an initial stellar population
much larger than the stellar content of today's GCs. This is generally attributed either to the
dynamic loss of a great fraction of the clusters' FG stars in the early evolutionary phases
\citep{dercole2008}, or to the formation of the GC within a much larger stellar environment, such as 
a dwarf galaxy, where the polluting matter is supplied by the surrounding stars \citep{bekki-norris2006}.

One of the important constraints for the progenitors was usually considered to be 
that their matter must have been processed
through the hot CNO cycle, and not, or only marginally, through the helium
burning phases, since the sum of CNO elements is the same within observational
errors in the ``normal" and
in the anomalous stars \citep{cohenmelendez2005, ivans1999}. 
If the polluting matter is identified with
the envelopes of asymptotic giant branch (AGB) stars \citep{ventura2001, ventura2002}, these
AGBs must be very massive, so that their evolution is only scarcely affected by
the third dredge up, that acts to increase the
surface abundance of the primary carbon formed in the helium intershell during the 
thermal pulses, and then partially mixed into the external envelope 
\citep[e.g.][]{ibenrenzini1983}. 
Otherwhise, pollution must come from a totally different kind of objects, such
as the envelopes of fast rotating massive stars during the core H--burning phase 
\citep{meynet2006,decressin2007}, 
and in this case we do not expect any C+N+O enhancement at all, as mentioned before.
  
Recently, the constancy of C+N+O has been challenged by the discovery that the cluster
NGC~1851 displays a double subgiant branch (SGB) \citep{milone2008}. 
No such splitting is present in the main sequence of NGC 1851, indicating a normal ---or close
to normal (see Sect.4)--- helium content 
for both SGBs. This has been interpreted by \cite{milone2008} 
as due to a $\sim$1Gyr age difference between the two SGB populations. This age difference is
much larger than generally believed possible for a GC \citep[e.g.][]{dercole2008}. 
An alternative interpretation
has been given by \cite{cassisi2008}: they were able to reproduce the splitting by assuming 
that stars in the faint SGB (fSGB) have a larger C+N+O abundance and similar age
of the bright SGB (bSGB). At the same time, \cite{yong2008a} have shown that 
this cluster harbors a star-to-star abundance variation in the s-process elements 
Zr and La and that the abundances of these elements were
correlated with the abundances of Na and Al, and anticorrelated with O. 
Furthermore, within the small sample of 8 giants examined, there
was a hint that the abundances of the s-process elements is bimodal.
Most recently, \cite{yong2009} have determined the C, N, and O abundances in four red giants,
selected to span the range of Na, Al and s--process abundances, and found that indeed the
sum of C+N+O exhibits a range of 0.6dex, a factor 4, and that the light elements abundances,
s--process and total C+N+O abundances are correlated.
These observations support indeed the AGB stars as progenitors of the stars populating the fSGB 
in NGC~1851, as suggested by \cite{cassisi2008}. 
Signs of a milder C+N+O enrichment in other clusters have been noticed by \cite{carretta2005} 
in NGC~6397 and NGC~6752, but NGC~1851 is to date the most prominent example 
\citep[see][for further examples]{piotto2009}.
As we have noticed, in most GCs the (possible) AGB progenitors of the SG
must be so massive that the chemistry of the ejecta is scarcely
affected by the third dredge up, with only a minor increase in the total
C+N+O abundance. But the effects of the third
dredge up increase with time: its efficiency increases when the evolving mass 
decreases \citep[e.g.][Table 1]{ventura2008a}. So the observability of a C+N+O increase depends on how long
the phase of star formation of the SG stars lasts. \cite{yong2009} propose 
that in NGC~1851 this phase lasts long enough that the effect of the third dredge up
is evident.

In this work, we consider the evolution of low mass stars evolving at GC ages, with 
the metallicity of NGC~1851, but considering the effect of an increased C+N+O abundance
on the evolution. \cite{cassisi2008} limited the analysis to a chemistry in which
the C+N+O content was approximately doubled with respect to the standard composition, while 
we consider C+N+O abundances increased by factors 2, 3 and 5. Selection of the individual
abundances for C, N and O is explained in Section 2. Inclusion of their effect is considered 
both in the low temperature and high temperature opacities and in the nuclear reaction network.
In Section 3 we present the results, and discuss the time evolution of selected
evolutionary tracks, the total mass lost along the RGB, 
and the relative location of the isochrones in the color magnitude diagram (CMD). 
In Section 4 we present simulations, in the CMD, of the main sequence and SGB for
NGC~1851, under several hypotheses concerning the different compositions necessary 
to explain the separate SGBs. In Section 5 we discuss the results. 

\section{Input physics and model computation}

All the evolutions presented in this work have been calculated by
means of the ATON code for stellar evolution, with the numerical 
structure described in details in Ventura et al. (1998). 
Tables of the equation of state are generated in the (gas) pressure-temperature
plane, according to the OPAL EOS of the Livermore group \citep[see OPAL webpage, 
last update in February 2006,][]{rogers1996}, 
replaced in the pressure ionization regime by the EOS by Saumon, Chabrier \& Van Horn (1995),
and extended to the high-density, high-temperature domain according to the
treatment by Stoltzmann \& Bl\"ocker (2000).  

\subsection{Standard and non--standard opacities}
For the ``standard" models we adopt the latest opacities by Ferguson et al. (2005) 
at temperatures lower than 10000 K and the OPAL opacities in the version 
documented by Iglesias 
\& Rogers (1996). The mixture adopted is alpha-enhanced, with $[\alpha$/Fe$]=0.4$
(Grevesse \& Sauval 1998). 
Electron conduction opacities were  taken  from the WEB site of Potekhin 
(see the web page http://www.ioffe.rssi.ru/astro/conduct/ dated 2006)
and correspond to the \cite{potekhin1999} treatment. The electron opacities
are harmonically added to the radiative opacities.
For this project we select three different mixtures of elements, having the
C, N and O abundances varied with respect to the standard mixture\footnote{We do not
include variation in the sodium abundance, as done by \cite{cassisi2008}, considering
that its total abundance is in any case very low, and its contribution both to
energy generation and opacity is negligeable.}. For
the abundances of C, N and O we adopt the values 
in Table 2 in \cite{ventura2008a}, corresponding to the 
yields of the models with 5, 4.5 and 4\msun\ and metallicity Z=10$^{-3}$. These abundances are reported 
in Table 1.\footnote{The choice of abundances has been motivated by the naive hypothesis that
the SG may directly be formed by the ejecta of the FG, and that the \cite{ventura2008a} yields represent the
SG composition. In fact, in Sect.~5, we will reinterpret the results in the light of a dilution model.
Notice however that direct formation from the FG ejecta does not necessarily imply an anomalous IMF peaked at the
masses whose yields are adopted, if we allow for ``self--enrichment" from a much wider cluster environment, as 
discussed in the Introduction.}
As we see, in the 5\msun\ ejecta Oxygen is depleted by 0.46~dex (the original [O/Fe] is +0.4 in
the standard mixture), Carbon is enhanced by 0.13~dex and Nitrogen is enhanced by 1.7~dex. The total
C+N+O is a factor 2.1 larger than in the standard mixture. The 4.5\msun\ total CNO 
content is 3.1 times the standard one, and the 4\msun\ one is a factor 4.9 larger. 
We remarked that the 4 red giants of NGC~1851 examined by
\cite{yong2009} show a maximum variation of $\sim$4 in C+N+O. Individually, carbon and oxygen vary up to 
about a factor 3 and nitrogen by a factor 7. With our present choices, we are then 
enhancing the nitrogen increase with respect to these few observational data. 
More observations and new models may be useful in future analyses. 
The total "metallicity" Z in mass fraction for the three  CNO--enhanced mixtures is  
listed in Table 1. 
%
\begin{table*}
\caption{Chemistry of the models}             
\label{chem}      
\centering          
\begin{tabular}{c c c c c c c c c}     
\hline\hline       
Name & Z & total CNO & [C/Fe] & [N/Fe] & [O/Fe] &  Y & M/M$_\odot$(AGB) & age (Myr)  \\ 
\hline            
CNOx1 (standard) & 0.00100 & 1 & 0.00  & 0.00 &  0.40  &0.240  & -&  -     \\
CNOx2    & 0.00185 & 2.1 &    0.13  & 1.70 & -0.06 &  0.324 & 5.0  & 103 \\
CNOx3    & 0.00235 & 3.1 &    0.12  & 1.89 &  0.19 &  0.310 & 4.5 &  128  \\
CNOx5    & 0.00350 & 4.9 &    0.14  & 2.02 &  0.44 &  0.281 &4.0 &  166  \\
\hline 
\hline
\end{tabular}
\end{table*}
The radiative opacities for these CNO enhanced mixtures have been computed
on purpose for this work. 
For temperatures above 10000~K we utilized the online computations from the OPAL 
group found at http://physci.llnl.gov/Research/OPAL/opal.html; 
for lower temperatures we computed opacities with the Ferguson et al. (2005) 
code for the Z mixtures 
listed in table~1. Figure \ref{opacity} shows the differences among the opacities
of the various mixtures in the high-- and low--temperature regime.
In the upper panel we see that the largest differences are found in the
ioniziation region of the CNO elements, around $\log T\sim 6$.
The lower panel of Fig.~1 shows the low temperature opacities for the three 
enhanced computations. Above $\log T\sim$3.5 few differences are noted as the opacity 
is dominated by hydrogen.  
However, at lower temperatures the opacity significantly deviates for each run from 
the standard.  This deviation is due to more and more O available (due to the CNO 
enhancements) for the formation of molecular water at these temperatures.
The low temperature opacities do not affect the structure of the models
we are considering, while at the larger temperatures of the stellar interiors 
some differences appear. In addition, the CNO total abundance affects 
the time evolution as soon 
as the stars begin evolving at the turnoff, where the CN cycle becomes important.

\subsection{Color --\teff\ conversions}

Our aim is to compare the theoretical predictions with the data by 
\cite{milone2008} for NGC~1851, obtained in the HST Advanced Camera for Surveys 
(ACS) filters F814W and F606W. Therefore we convert our L,\teff\ values into 
these magnitudes by using the transformations for ACS bands by Bedin et al. 
(2005), based on the solar scaled models by \cite{cassisi2004}. We note that the
$\alpha$--enhanced versus solar--scaled transformations, for the metallicity 
and colors we are using, do not differ for \teff$\simgt$5000K. On the other hand,
we are also dealing with CNO enhanced mixtures, and in principle the use of the same
transformations could not be adequate. However, we refer
to \cite{pietrinferni2009}, who widely discuss this issue, and conclude that,
in broadband filters not bluer than the Johnson V band, the effect of this
inconsistency should be very small, since at least the color -–
\teff\ transformations are largely independent on the metals
and their distribution \citep[see][]{alonso1996,alonso1999,cassisi2004}.
This statement luckily applies to our case, as we are dealing with red and near infrared bands.   
Possible differences in the conversions due to the enhanced CNO may of course change the
quantitative conclusions of our analysis, so that we raise the problem of the computation of proper
color--\teff\ conversions for CNO enhanced mixtures.

\begin{figure}
\includegraphics[width=8cm]{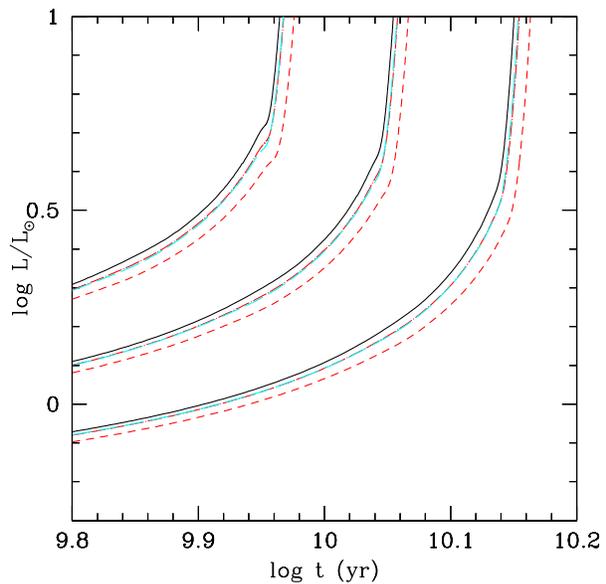}
\caption{Luminosity versus time evolution for the masses 0.8, 0.85 and 0.9\Msun (from right to left) as 
a function of the CNO enhancement. The full (black) lines are the CNOx1 evolution.
The CNOx2, CNOx3 and CNOx5 tracks are located progressively at smaller luminosity.}
\label{f1}
\end{figure}

\begin{figure}
\includegraphics[width=8cm]{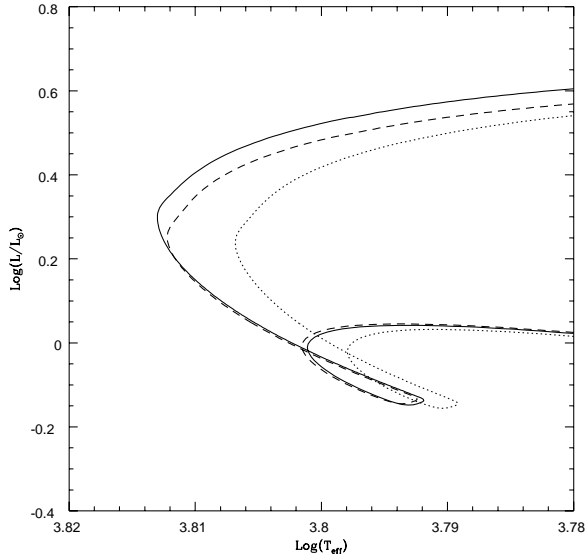}
\caption{Comparison between the HR diagram of 
the standard CNOx1 track of 0.85\Msun (full line) and of the CNOx3 track (dotted line), with
the evolution of the same mass, obtained by assuming the opacities of the
standard track, but the CNO abundances of the CNOx3 track (dashed line). 
}
\label{f1franca}
\end{figure}

\begin{figure}
\includegraphics[width=8cm]{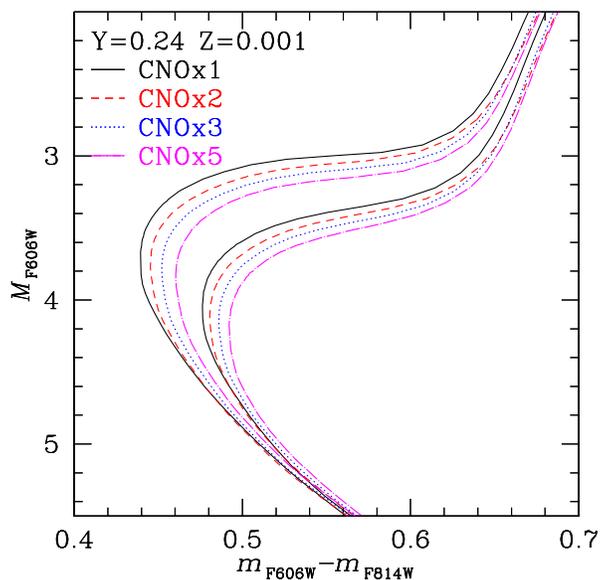}
\caption{Isochrones in the plane $M_{\rm F606W}$ vs. the color $M_{\rm F606W}-M_{\rm F814W}$. 
The top group of lines refers to 
an age of 9Gyr, the lower set are for an age of 13Gyr. The isochrone luminosity decreases
as a function of the CNO enhancement.}
\label{f2}
\end{figure}

\begin{figure}
\includegraphics[width=8cm]{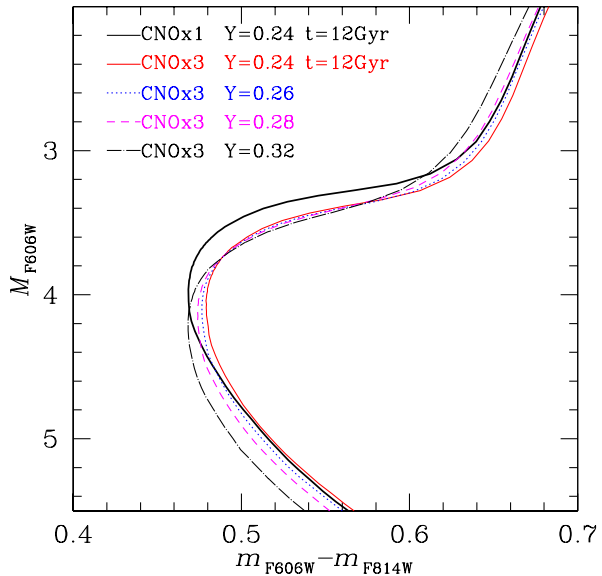}
\caption{Isochrones in the plane $M_{\rm F606W}$ vs.  $m_{\rm F606W}-m_{\rm F814W}$. 
The standard Y=0.24 CNOx1 
t=12Gyr isochrone is the upper line. The lower lines are for the same age and 
different helium content. The SGB luminosity is only slightly altered for Y=0.32.}
\label{f3}
\end{figure}

\begin{figure}
\includegraphics[width=8cm]{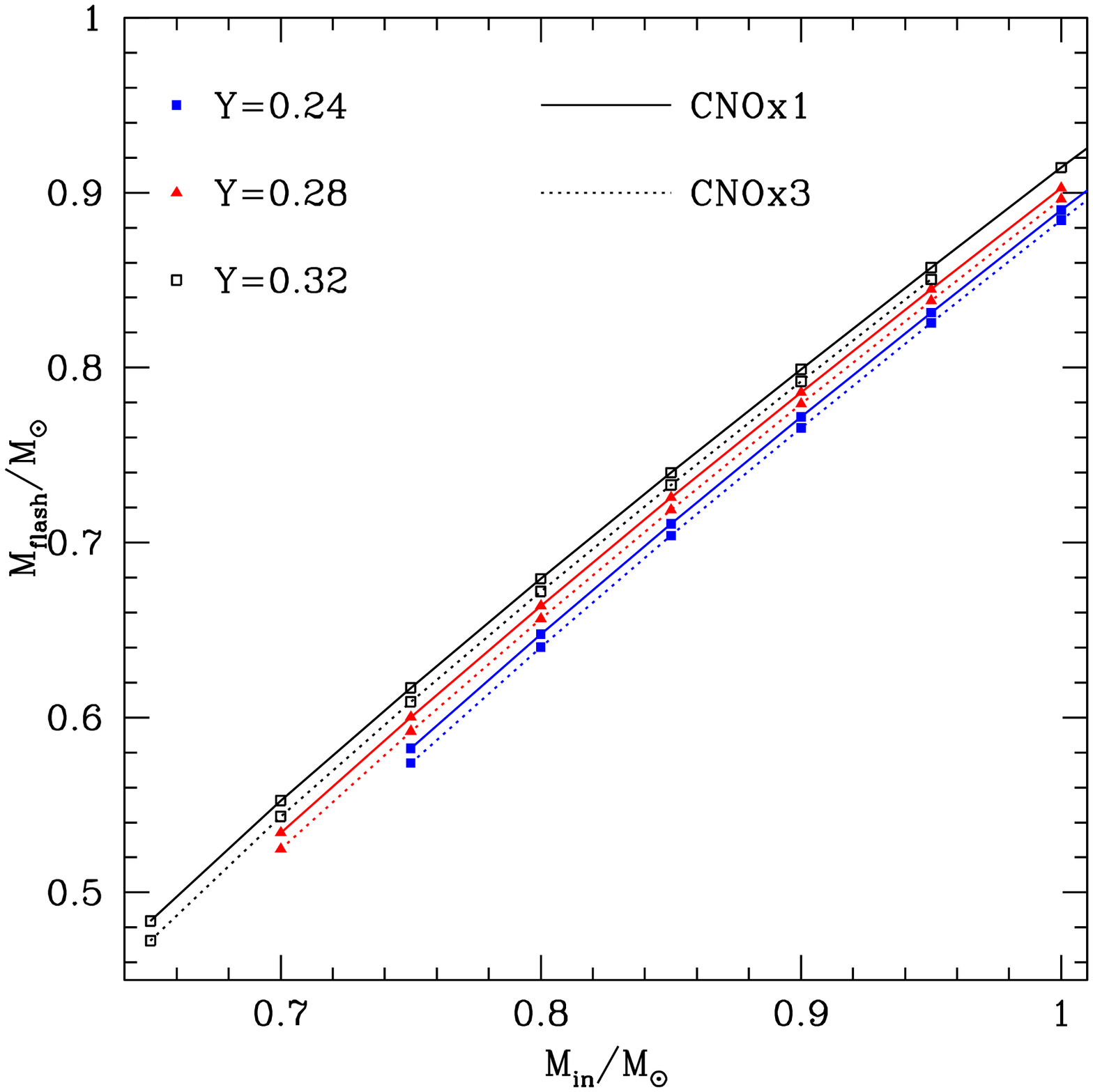}
\includegraphics[width=8cm]{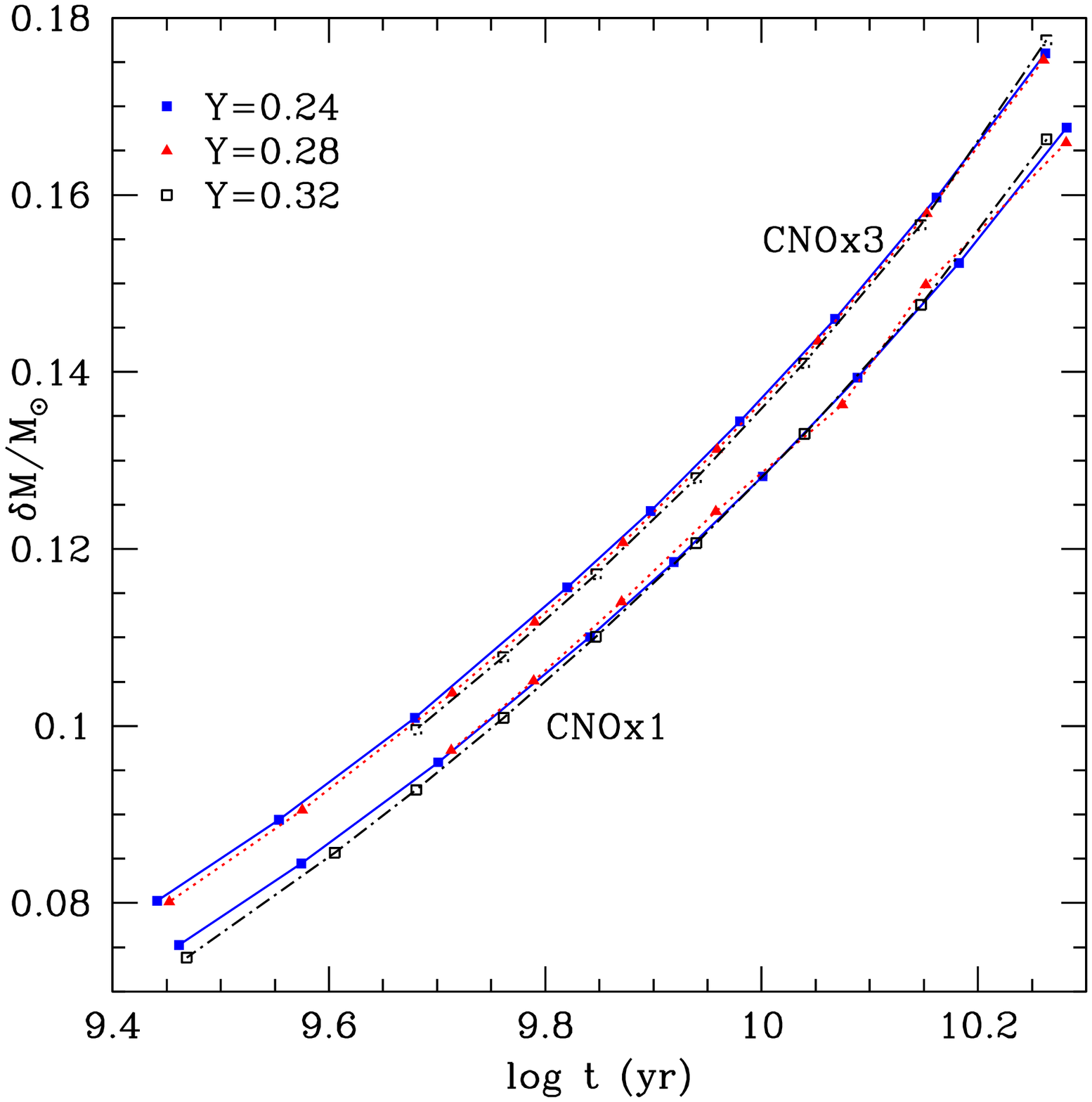}
\caption{The top panel shows the mass at the helium flash, M$_{\rm flash}$ as a function of the total stellar mass,
for the three labelled helium contents and for the standard tracks. The dotted lines show the same
for the CNOx3 tracks. Although for a fixed mass there is a dependence of M$_{\rm flash}$  
on Y and CNO, the bottom panel shows that the mass lost along the RGB ($\delta$M)  
practically does not depend on the helium content at
a fixed epoch. On the contrary, at a given age, $\delta$M is $\sim 0.01$\Msun\ 
larger for the CNO enhanced
isochrones. We plot in the figure only the CNOx1 and the CNOx3 cases, 
the other two cases are practically superimposed to the CNOx3 lines.
}
      \label{fmassloss}%
\end{figure}

\subsection{Synthetic CMD Simulations}
For a better comparison of the models with the data, we use simulations for the main sequence and 
subgiant branch(es). These are obtained by extracting random each star location along a specified
isochrone, according to a choice of the initial mass function (IMF). We assume a power
law IMF with exponent --1.5 (where Salpeter's is --2.3). The exact choice 
of the IMF is inconsequential, as the 
comparison of the SGBs involves a very small mass interval (see Sect.~4).
Gaussian errors 
consistent with the observational errors on colors and magnitudes close to the turnoff 
are attached to each extraction. For the double stellar population of NGC~1851, we use a standard isochrone
of CNOx1 for 55\% of the sample, while the rest is extracted from the isochrone having the same age and
enhanced CNO. These percentages are based on the observational analysis by \cite{milone2009}, and are 
confirmed in the analysis of the subgiant branches made in Sect.~4, describing the
comparison between the simulations and data.

\section{Model results}
\subsection{The isochrone location: dating GCs with the same metallicity and different CNO}
Figure \ref{f1} shows the luminosity versus time evolution of masses 0.8, 0.85 and 0.9\Msun\ for different
CNO enhancement. The CNOx1 standard tracks have the fastest evolution, because the CNO enhanced tracks 
lie at progressively lower main sequence luminosity. 
In order to understand better the different roles of the opacities and nuclear reaction rates, we show in Figure
\ref{f1franca} the HR diagram location of a track of 0.85\Msun, having the standard CNOx1 opacities, but
CNOx3 abundances (in particular the CNOx3 abundances are used in the nuclear network). 
We see that the main sequence evolution is quite similar to that
of the standard CNOx1 track, as burning occurs mainly through the proton--proton chain. The main sequence  
of the track is just slightly more luminous than for the standard CNOx1 track, and the evolutionary
times are consequently barely shorter. The turnoff,
however, occurs at a smaller luminosity, as in that case the CNO cycle dominates. Therefore, 
the lowering of the turnoff luminosity is mainly due to the effect of the abundances on the final phases 
main sequence burning, while the lenghtening of the evolutionary times is mainly due to the smaller
main sequence luminosity of the models with enhanced CNO. 

Globally, the isochrones show a monotonic shift in luminosity and color with increasing 
CNO. Figure \ref{f2} shows that the subgiant location
for the same age is fainter for higher CNO, so that these models can in principle reproduce the
splitting of the SGBs in NGC 1851, if the two sequences have different CNO content.
Figure \ref{f3} shows the comparison between the standard SGB for Y=0.24 and the SGBs of
the models CNOx3 with different helium contents (Y=0.24, 0.26, 0.28 and 0.32). We see that
the SGB shift does not depend on the helium content, at least for this metallicity, while the
main sequence location and the red giants become bluer when Y increases (see also Sect.~5). 

Considering Fig. \ref{f2}, we observe that any age determination for a GC can not ignore what is the
total C+N+O abundance in its stars. While this is a well known theoretical aspect of evolution
\citep[e.g.][]{simoda1970, bazzano1982}, it is worth recalling it in this context, now that we have
interesting evidence of its possible role. In fact, the interpretation of NGC~1851 data shows 
this  \citep{cassisi2008}, and we also have other cases where it may be of relevance \citep{piotto2009}.
The total CNO abundance in the giants of M4 (NGC 6121) is larger than the total CNO of those
stars, in NGC~1851, that can be considered ``normal" for what concerns the abundances of 
CNO, s-process elements, Na and Al \citep[see Figure 3 in][]{yong2009}. 
Therefore, should we compare
these two clusters, their relative age determination should include the dependence on the CNO abundance.
In this context, it is worth recalling the large spread in N abundance in all
the clusters surveyed \citep{grundahl1999}, together with the dichotomy
between CN strong and CN weak stars in so many GCs. The giants in NGC
6752 present a star-to-star abundance variation in N of 1.95dex \citep{yong2008b}, 
and a similar spread is present also in the main
sequence \citep{pasquini2008,carretta2005}. Variations in C
abundance may compensate, but some spread in luminosity at the turn-off of
this cluster should be present \citep{bazzano1982}. 

\subsection{The mass lost along the red giant branch}
All models are evolved by assuming a mass loss rate following Reimers' (1977) prescription
$ \dot M_R=4 \cdot 10^{-13} \eta_R {LT\over M} $
where the parameter $\eta_R$\ has been fixed to 0.3 for all the computations. This allows us to
find the dependence of the mass lost in the giant phase on the helium and CNO content of the tracks. Of course the 
results are of some significance only if Reimers' rate is actually a good description of the mass loss.
In a relative sense, the results may be similar for other mass loss rates, if they depend explicitly only on the
stellar parameters luminosity and gravity, and do not depend, e.g., on the CNO content.
Some results are shown in Fig. \ref{fmassloss}. In the top panel we see that mass at the helium flash,
for a fixed initial mass, increases with Y 
and decreases with the CNO enhancement. We plot the CNOx1 and CNOx3 case,
but the results for the other CNO rich models are similar. However, 
what matters is the mass lost at fixed age,
shown in the bottom panel of Fig. \ref{fmassloss}. This actually does not depend at all on Y 
\citep[see also][]{dantonacaloi2008},
while, for a fixed age, all the CNO enhanced models predict $\sim$0.01\Msun\ 
more mass loss, as shown for the case of the CNOx3 models.
These indications are important to predict the horizontal branch (HB) location 
of stars with different CNO and/or helium
content. \cite{cassisi2008} have shown that the \teff\ location of HB models 
is affected by their enhanced CNO, due to the different 
relative efficiency of the shell--hydrogen-- and core--helium-- burning 
(see their Figure 1, but see also \cite{castto1977}).

\section{Comparisons}
\begin{figure}
\includegraphics[width=7cm]{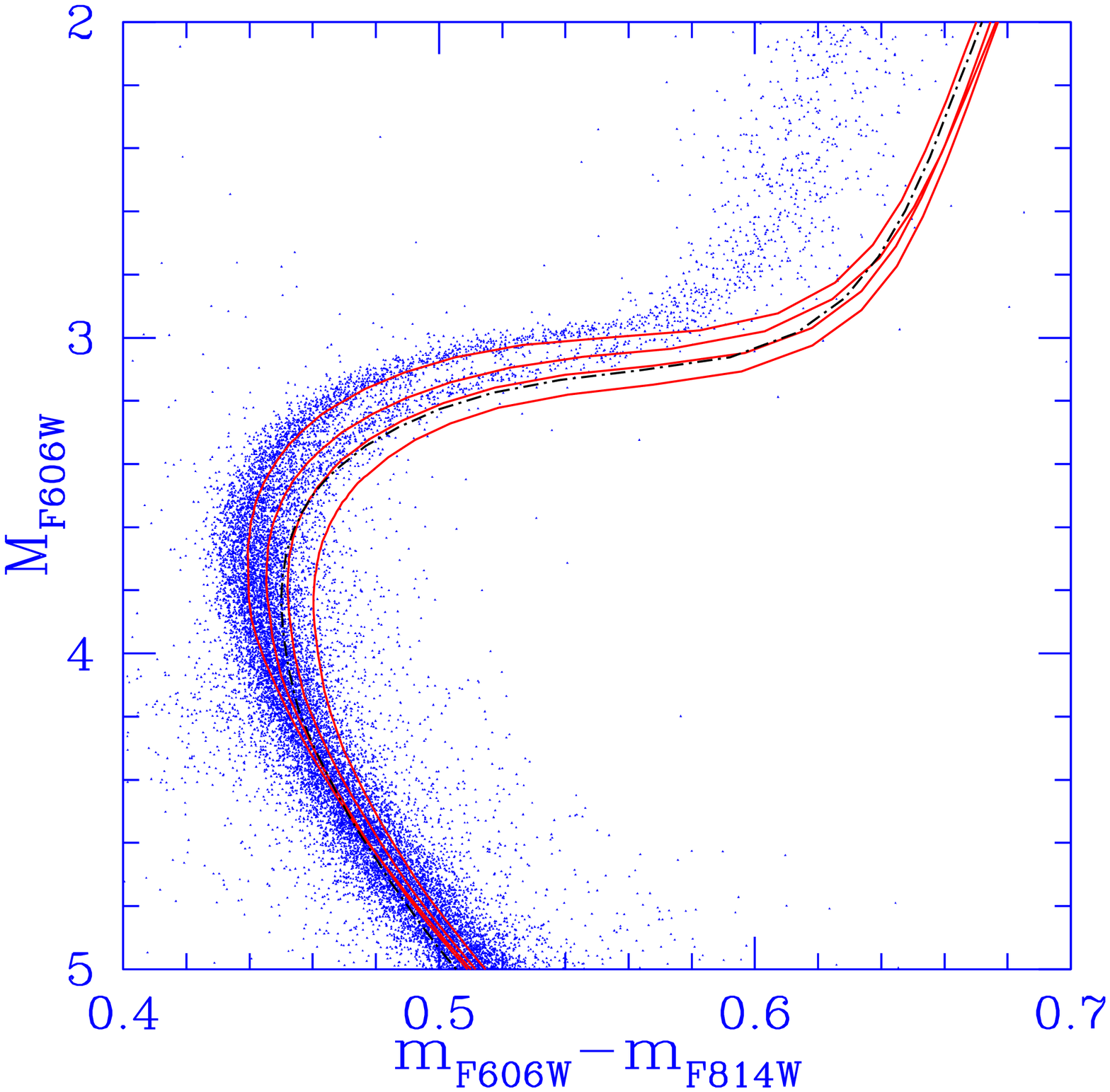}
\includegraphics[width=7cm]{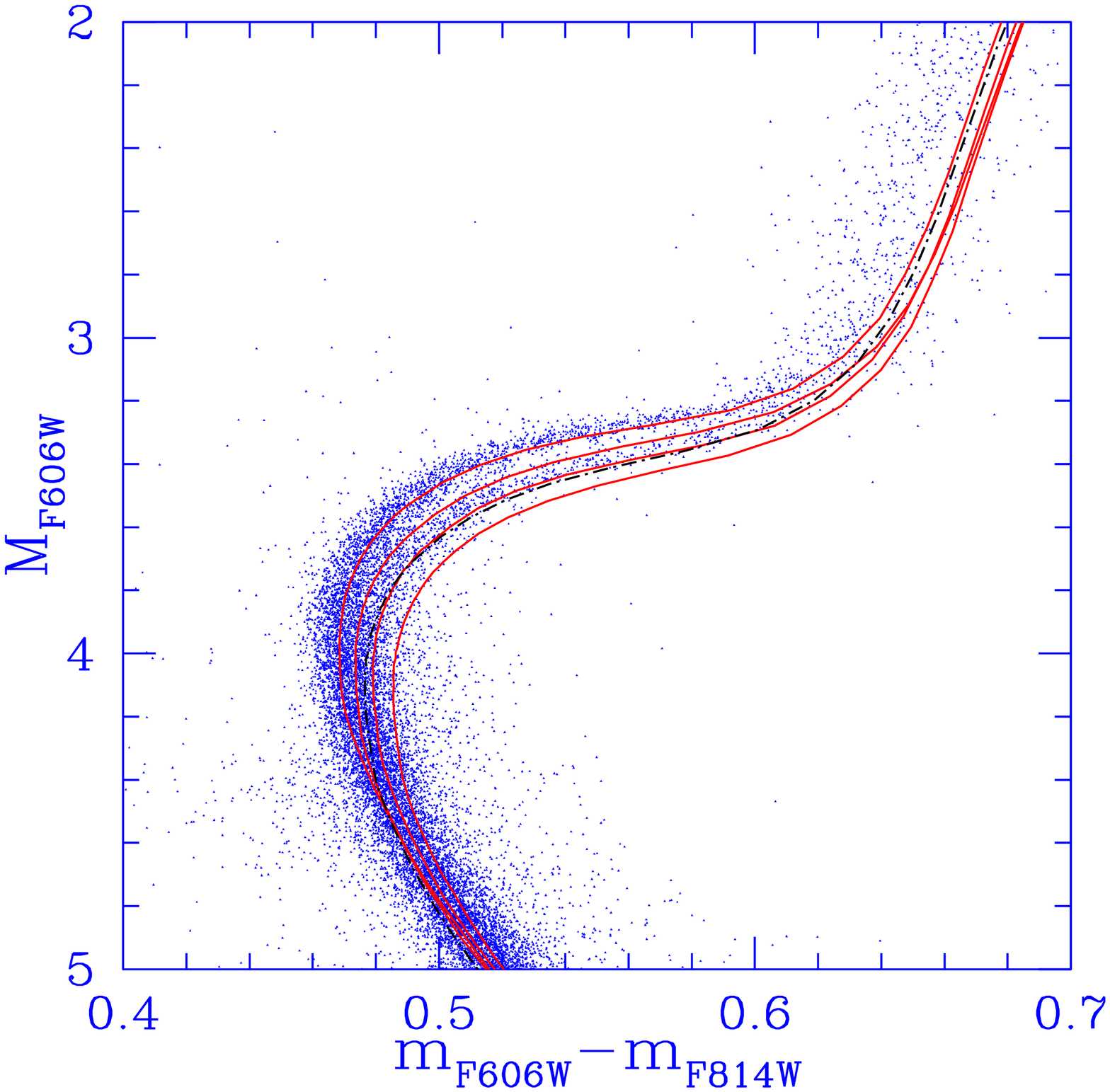}
\caption{Comparison of isochrones of ages 9~Gyr (top panel) and 12~Gyr (bottom panel)
with the HR diagram of NGC 1851. We apply to the data
a distance modulus and dereddening in order to fit the ``standard" top isochrone to
the bSGB. The dashed line (black) is the CNOx3 isochrone of Y=0.26.
Distance moduli and reddenings respectively: ($m_{\rm F606W}$--$M_{\rm F606W}$)=
15.72, $\delta$color=0.05mag for the top panel (9Gyr isochrones) and 15.44mag, 
0.02mag for the bottom panel (12Gyr isochrones). The continuous (red)
lines, starting from the top to
the bottom one, are the isochrones for Y=0.24 and respctively CNOx1, CNOx2, CNOx3 
and CNOx5.}
      \label{f7}%
\end{figure}

Figure \ref{f7} shows the HR diagram of NGC~1851 based on 5 images of 350s in F606W and
5 images of 350s in F814W from the ACS survey  by \cite{anderson2008}. Thes exposure are 
saturated at the basis of the RGB, so the photometry of the RGB comes from two low exposure images 
(one 20s image in F606W and one 20s image in F814W) and has larger errors\footnote{
Notice that the RGB of this cluster is broadened --or even split-- if we observe it in
the U, B, or other filters in which the CN--bands are present \citep[e.g.][]{lee2009}, but 
the reason why the RGB is so broad in these observations is simply due to the photometric error.}.
The data in Fig.\ref{f7} are compared with the isochrones. Distance and
reddening are chosen to fit isochrones of ages respectively 9 and 12Gyr in the two panels.
We see that the giant branch location agrees better with the 12Gyr isochrones, but it is
difficult to attribute to this feature a predictive value for the age, as the giant branch
location depends on the adopted convection model.

\begin{figure}
\includegraphics[width=7.0cm]{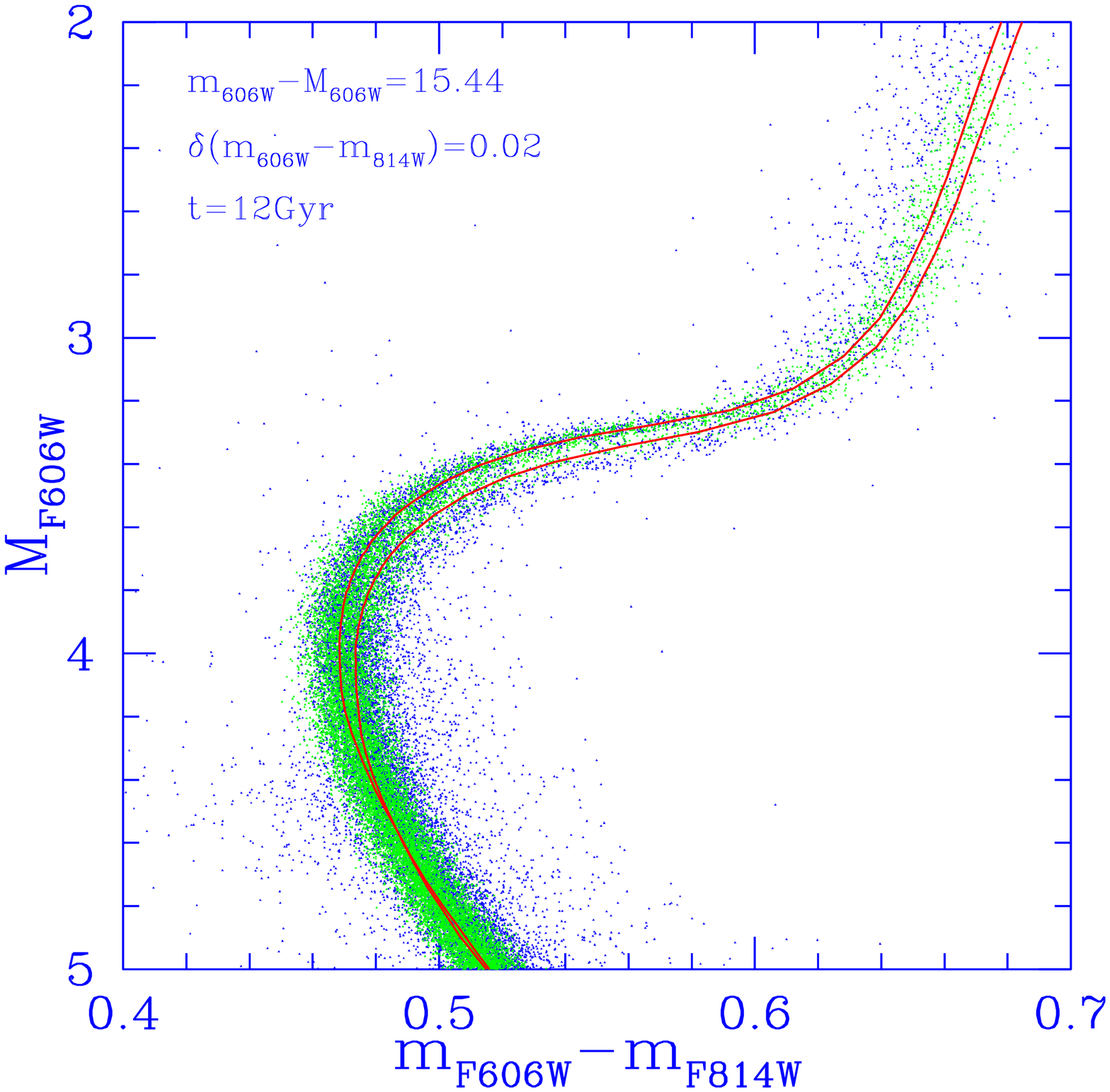}
\includegraphics[width=7.0cm]{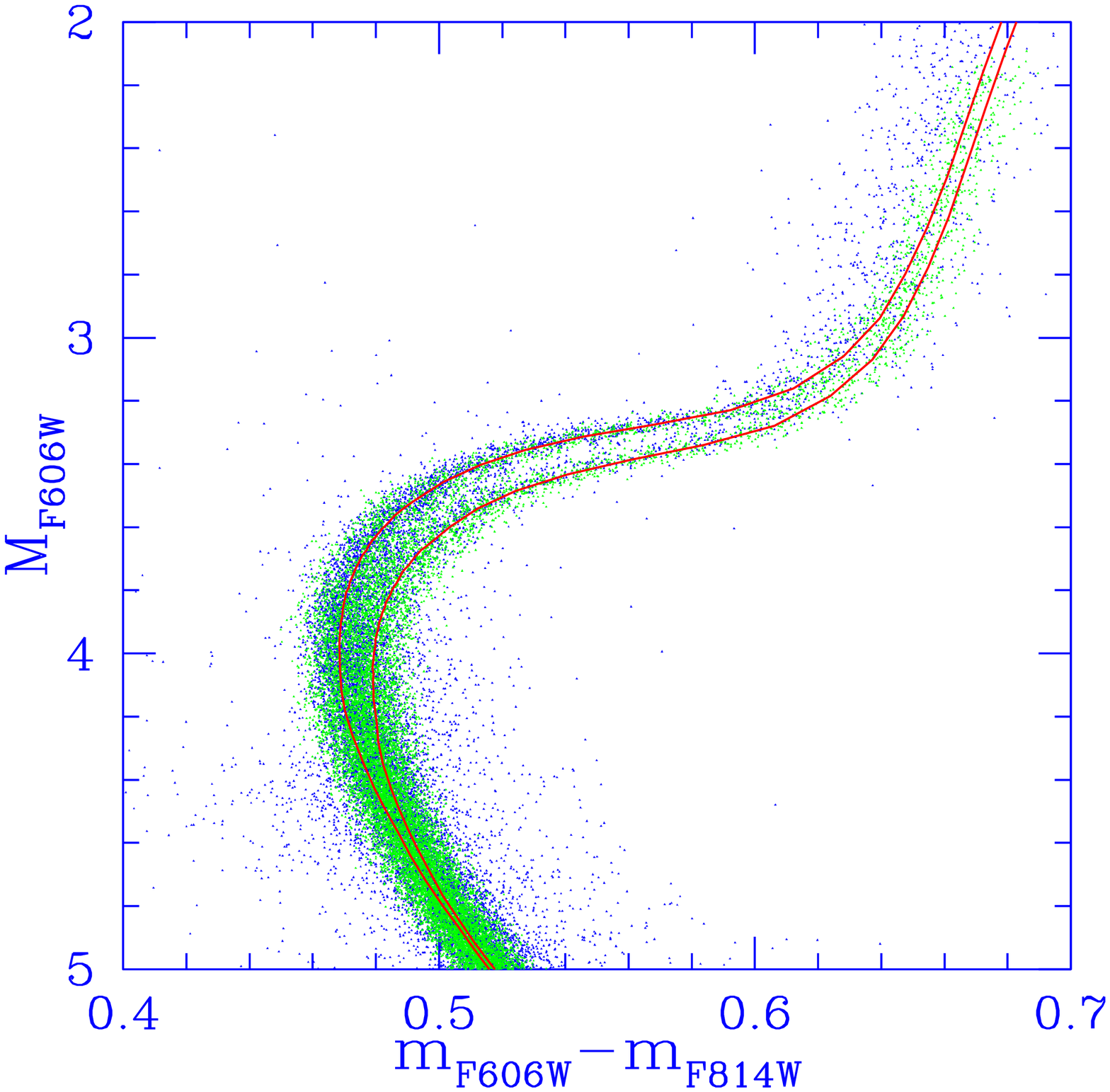}
\includegraphics[width=7.0cm]{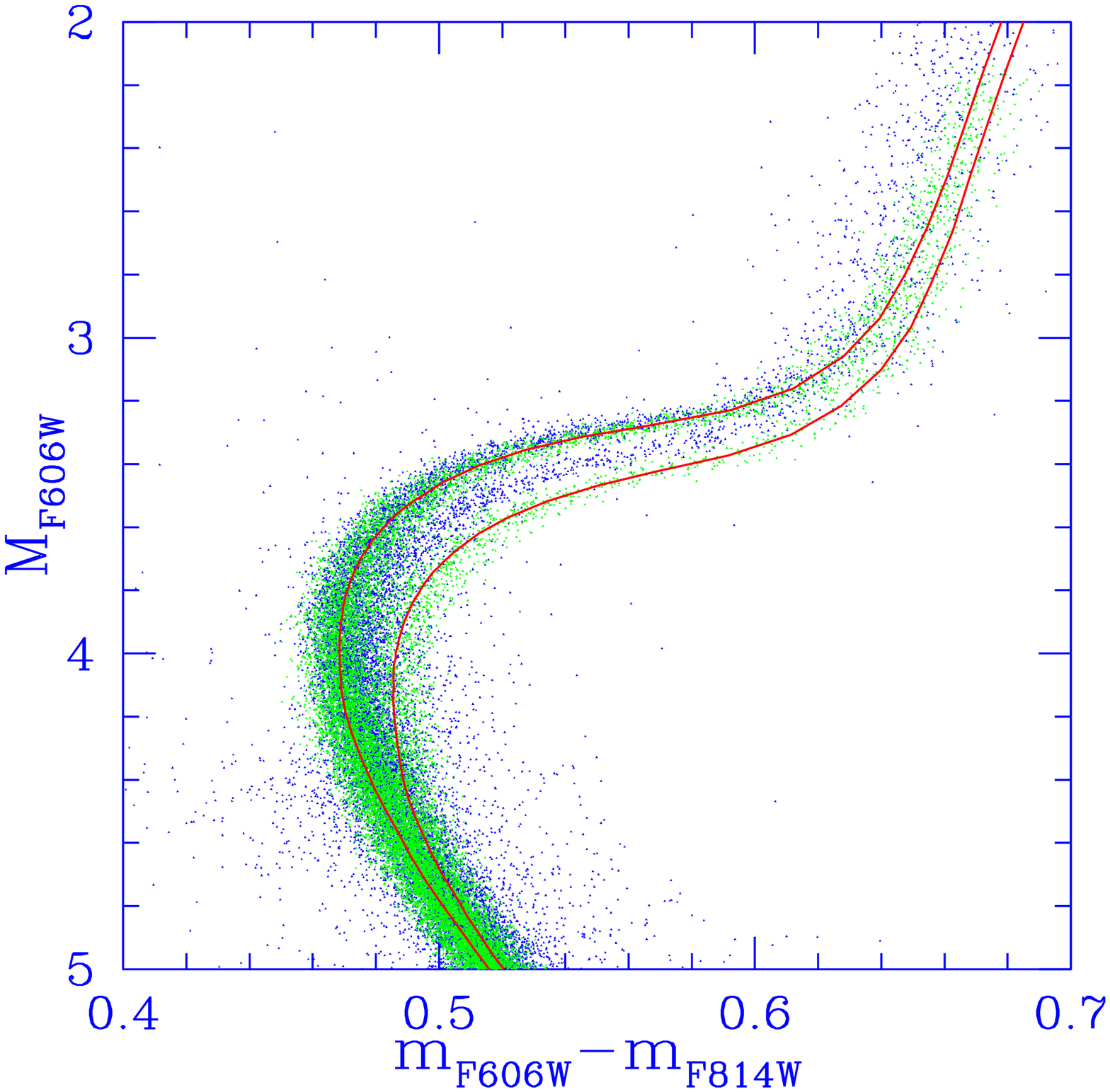}
  \caption{Comparison of NGC~1851 data (blue dots) with simulations (green) for an age of
  12 Gyr. Distance modulus and assumed reddening are labelled. The bSGB is simulated with 
  a population having normal (CNOx1) CNO, the fSGB corresponds to a population with
  CNOx2 (top), CNOx3 (middle) and CNOx5 (bottom panel). The corresponding isochrones
  of 12~Gyr are also shown. The helium abundance is assumed to be Y=0.24 for both 
  populations.}
  \label{f8}
\end{figure}

This comparison immediately shows that the fSGB in the cluster lies in between the isochrones
corresponding to CNOx2 and CNOx3.
This finding is strengthened when we make use of the simulations built up as described in
Sect.~2.3. Figure \ref{f8} compares the 
observed HR diagram with simulations for an age of 12 Gyr. The bSGB is reproduced with a standard
CNOx1 population, while the fSGB is reproduced assuming a second population with CNOx2
(top), CNOx3 (middle) and CNOx5 (bottom panel).
It is evident, looking at the bottom panel of Fig. \ref{f8}, that the CNOx5 isochrones are
not adequate to describe the splitting of the SGBs: such a CNO enhancement would produce a 
splitting much larger than observed. The CNOx2 isochrones produce a very small splitting, 
the CNOx3 isochrones are the closest to the observations.

\begin{figure*}
\includegraphics[width=14cm]{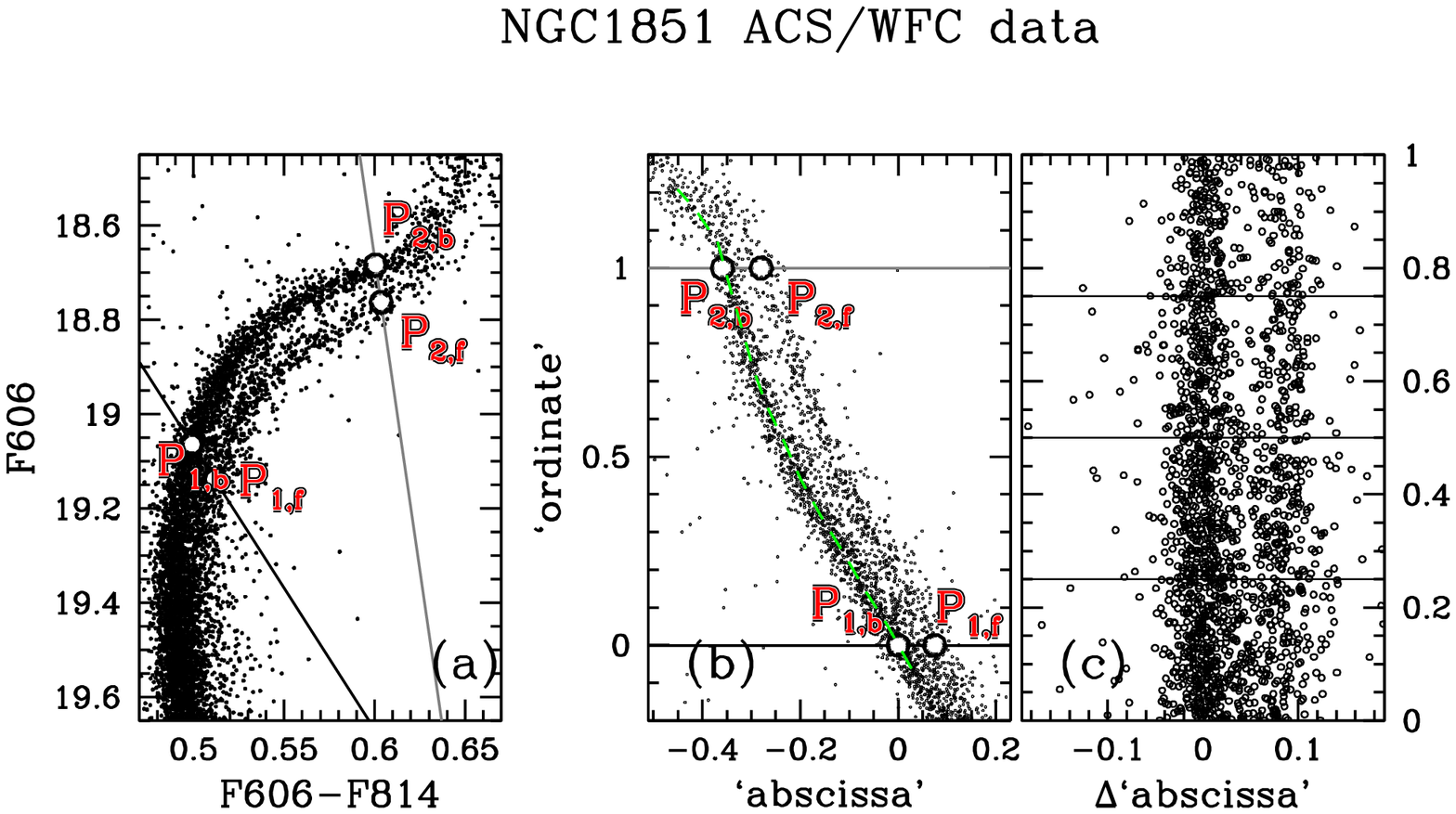}
\caption{This  figure  illustrates   the  procedure  for the comparison of the  observed data and the simulation.
	Panel  (a)  is  a  zoom  of  the  observed ACS/WFC HR diagram.
        The two  lines delimit the portion of the HR diagram where the split is
        most evident. Only stars
        from this region are used to measure the population ratio.  In
        Panel (b) we have linearly transformed the reference frame of Panel (a).
        The green dashed line is  the fiducial of the region around the
        bSGB.  In Panel  (c) we plotted stars between  the two lines
        but after the subtraction of  the `abscissa'. }
\label{f9}
\end{figure*}

For a quantitative comparison of observed and simulated HR diagrams
we go through  a procedure already used by  \cite{milone2009}.
Fig.~\ref{f9} illustrates this three-step procedure for the observed ACS/WFC data.
We  selected by hand two points  on the fSGB ($P_{1,f}$,$P_{2,f}$) and
two points  on   the   bSGB  ($P_{1,b}$,$P_{2,B}$) with  the  aim   of
delimiting the  SGB region where  the split  is most  evident.   These
points define the two lines in panel  (a), and only stars contained in
the region between these lines were used in the following analysis.
In panel (b)  we have transformed the  CMD  linearly into  a reference
frame where:  the origin corresponds to  $P_{1,b}$; $P_{1,f}$ has unit
abscissa,  and both $P_{2,b}$ and  $P_{2,f}$ have  unit ordinate.  For
convenience,     in the  following,  we   indicate   as `abscissa' and
`ordinate' the abscissa and the  ordinate of this reference frame. The
dashed green line is  the fiducial of the bSGB.  We drew it by marking
several points on the  bSGB, and interpolating  a line through them by
means of a spline fit.
In panel (c) we have  calculated the difference between the `abscissa'
of   each star and  the `abscissa'    of the  fiducial  line ($\Delta$
`abscissa').

\begin{figure*}
\includegraphics[width=14cm]{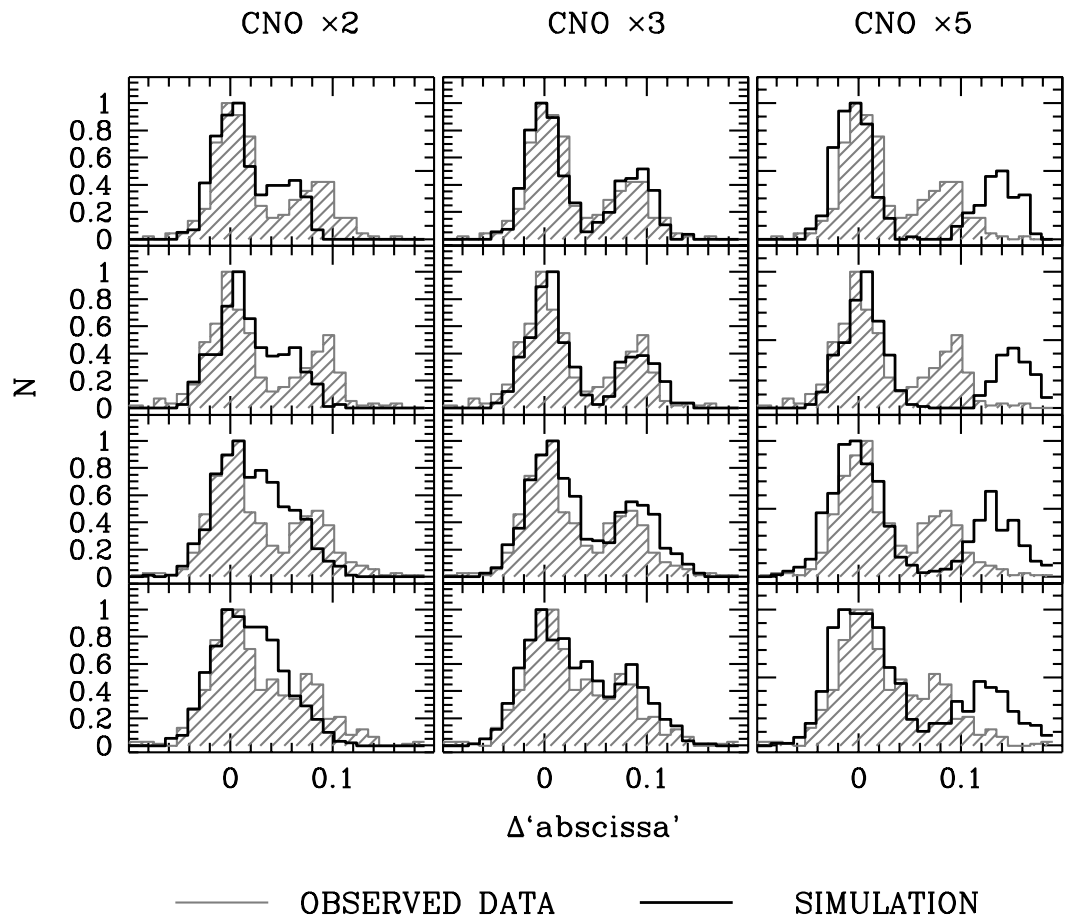}
\caption{ 	The  histograms  show the
        $\Delta$`abscissa'    distribution     for    stars    in    four
        $\Delta$`ordinate'   bins.   We compare   the   distribution of  the
observed    data  (grey    histograms)  and   the  simulations  (black
histograms).
}
\label{f10}
\end{figure*}

The histograms in Fig.~\ref{f10}  are the normalized distributions in $\Delta$
`abscissa'  for stars in four $\Delta$ `ordinate' intervals.  In the
three panels  of   this figure we compare   the   distribution of  the
observed    data  (grey histograms)  and   the  simulations  (black
histograms). The bSGB corresponds to the population  with normal CNO, the
fainter SGB corresponds to the CNO $\times$ 2  (left),
CNO $\times$ 3 (middle), and CNO $\times$ 5 (right).

If we use slightly helium enriched isochrones (up to Y$\sim$0.28) for the simulation of the
fSGB, the results do not change for what concerns the SGB splitting (see Fig. 4).
Notice however that the HB observations pose a constraint on the maximum helium
abundance of the SG in this cluster: \cite{salaris2008} were not able to reproduce the HB morphology 
attributing to the whole blue HB side (probably corresponding to the same population of
the fSGB) with models having helium content as high as Y=0.28 (and for larger Y the situation
is worse). From the width of the MS, however, we can derive an independent upper limit to the helium
abundance of the SG stars, by comparing it with models having different Y. 

\begin{figure*}
\includegraphics[width=14cm]{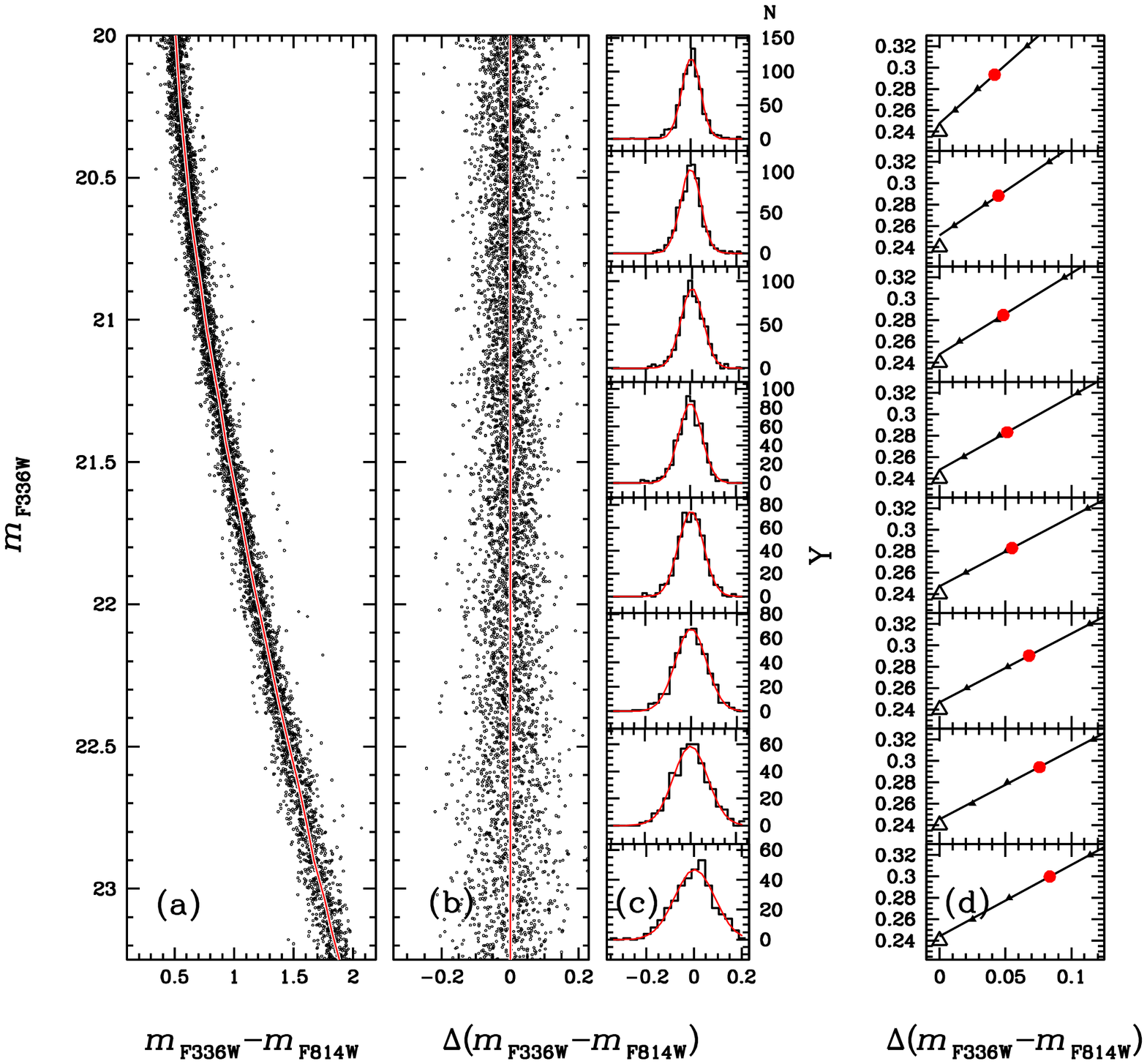}
\caption{Panel $a$ shows  the ${\it m}_{\rm F336W}$  vs.  ${\it m}_{\rm F336W}-
{\it m}_{\rm F814W}$ CMD  already presented in  Milone et al. \ (2008)
with the fiducial line overimposed.
Panel $b$ shows the same CMD, after subtracting from each star the
color  of the fiducial line  appropriate for its ${\it m}_{\rm F336W}$
magnitude.
The  histograms of the color distribution  in eight magnitude bins are
plotted in panel $c$.
In panel $d$ we show the helium content Y as a function of the color distribution in each
bin.   The  three  black triangles indicate the MS spread predicted by teoretical
models for a SG with CNOx3 and Y=0.26, 0.28, and 0.32, and are used to derive the
straight line by means of a least squares fit. The open triangles
corresponds to the case where the SG has CNOx1 and Y=0.24. Red circles indicate the
Y value corresponding to the observed MS color spread.
}
\label{f11}
\end{figure*}

Figure \ref{f11} shows 
such a comparison for eight magnitude bins of the MS in the color ${\it m}_{\rm F336W}-
{\it m}_{\rm F814W}$. From the mean of the  eight Y extimates, we derive  an upper limit for
the He spread in NGC~1851. We find $\Delta Y$=$0.043 \pm 0.003$. If the FG has Y=0.24,
the maximum possible helium abundance value is Y=0.290$\pm 0.003$.

\section{Discussion}

The computed models and the comparison with the data show that the CNO ehnancement required
to explain the double SGB in NGC~1851 is of about a factor three. This confirms
and puts on a more quantitative basis \cite{cassisi2008} results. 
(In the following discussion, anyway, keep in mind that the quantitative conclusions 
of our analysis are based on the use of the same
color--\teff\ relations both for CNO normal and CNO enhanced mixtures).
So, in this cluster, we {\it need} CNO enhancement to fit the HR diagram features. If we add the
spectroscopic information from \cite{yong2008a} and \cite{yong2009}, on the presence
of s--process enhancement and of a global CNO spread among the cluster giants, 
we may conclude that massive AGB pollutors are the only reasonable candidates to form the SG stars in 
the fSGB in NGC~1851.

It is possible to better quantify this statement by looking at Table 1. 
We choose our compositions, 
for the elemental abundances and opacity computation, from the pure ejecta of our
AGB models of Z=10$^{-3}$ \citep{ventura2008a}. In the table we see that a factor three in CNO is
reached in the matter expelled by the 4.5\msun\ stars. 
If, however, the matter forming the SG stars 
comes from pure ejecta only, these stars should have a helium content of Y$\sim$0.31.
Indeed such a large helium content is to be excluded, based 1) on the main sequence, 
as it does not show the width required by such a large Y spread (see, Figure \ref{f3}
and the discussion in Sect. 4);
and 2) on the HB morphology, both because the HB lacks very hot stars, and because the blue 
side of the RR Lyr region is not particularly overluminous \citep[e.g.][]{salaris2008}.

A possible solution is that the AGB ejecta have been diluted with pristine matter
while forming the SG stars. If, e.g., the matter comes preferentially from the 4\msun, the ejecta
would have 5 times the standard CNO, as listed in Table 1. 
Dilution with 50\% of pristine matter gives the
required CNO enhancement by a factor 3. In this case, the starting helium abundance 
in the ejecta was Y=0.28. Diluting it with 50\% of matter at Y=0.24, the helium content of 
the SG stars have Y=0.26, a value not in contradiction with what required by 
the HB morphology. 

It remains to be understood why the SG took such a long time to be formed (according to
Table 1, 166Myr is the age at which the 4\msun\ has evolved and ejected its
CNO enriched matter into the cluster), while in most other clusters it seems to be formed much earlier,
within at most 100 Myr. In fact, Table 1 shows that
the masses contributing to the SG must be larger than $\sim$5\Msun, if the total CNO content
has to be kept within a factor two of the FG value \citep{cohenmelendez2005, ivans1999}. In
addition, the very high helium subpopulations found in a few very massive clusters
\citep{piotto2009} require that, in these clusters, part of the SG has been
formed by the ejecta of the most massive super--AGB stars \citep{pumo2008,dercole2008}, 
evolving at ages $\sim 30-40$Myr.

The necessity of a C+N+O enhancement in the SG of NGC~1851 is further established with our analysis,
following the \cite{cassisi2008} discussion, and has at least an initial observational basis in the
\cite{yong2009} data. The above detailed formation scenario for this cluster, however, relies
on the yields from the models by \cite{ventura2008a} that have been employed in this discussion. 
Different models, e.g. the set by \cite{karakas2007}, provide stronger effect of the third dredge up
at larger initial masses, mainly because these models adopt of a less efficient convection 
description \citep[see the discussion in][]{ventura2005a}. The \cite{karakas2007} yields
do not favour the AGB interpretation of observational data for most 
clusters, namely the quasi--constancy of C+N+O and the shape of the anticorrelation 
O--Na \citep{fenner2004}. A possible escape is that the whole range of pollutor masses for the SG 
is shifted to the super--AGB range, but, to our knowledge, no such models are available by now. 
However, yields close to those by
\cite{karakas2007} would change our conclusions on the timescale of formation of an SG, just because the
stronger effect of the third dredge up ---limited, anyway, to a factor $\sim$3 increase in
C+N+O--- would be obtained at larger evolving masses ($\sim 6$\Msun) and much shorter
evolutionary times.

\section{Conclusions}
We computed the evolution of standard $\alpha$--enhanced low mass stellar models 
of metallicity Z=10$^{-3}$, and of models with the same [Fe/H], in which the total
CNO abundance is larger by factors 2, 3 and 5, according to the elemental distribution
expected from the intermediate mass AGB ejecta that can be progenitors of SG stars
in Globular Clusters. Our aim is to derive quantitative information about the C+N+O difference
required to explain the splitting of the subgiant branch in the GC NGC~1851.
Comparison of the models with the data quantifies the necessity of C+N+O enhancement
in the fSGB. The amount of total CNO required is a factor of about three larger than the total
CNO adopted for the bSGB. We incidentally warn 
that a difference in total CNO between clusters having the
same metallicity may simulate a non negligible age difference.
We try to interpret the data on the basis of the massive AGB scenario for the formation of 
the fSGB, showing the presence of an SG. The ejecta of the AGBs
that formed the SG must have a total CNO abundance increased by a factor three at least.
Though, if these pure ejecta have formed the SG, its helium abundance would be as high
as Y=0.31, incompatible with both the MS color width and 
the morphology of the HB. We propose that the CNO enrichment in
the ejecta was a factor at least 5 larger than the initial CNO, and that the ejecta
have been diluted with pristine gas by 50\%. In this case, the abundance of helium in the
SG comes down to a much more reasonable value of Y=0.26 or less, but we must explain why
the SG formation was delayed to a total age of more than 150Myr in this cluster and not
in others.
Finally, we remember that the quantitative conclusions of our analysis are based on the use of the same
color--\teff\ relations both for CNO normal and CNO enhanced mixtures. Although this seems reasonable for
the red bands we are using \citep{pietrinferni2009}, the possibility that appropriate color--\teff\ relations,
when available, may modify the CNO enahncement required to fit the two SGBs in this cluster must be kept 
in mind.

\section{Acknowledgments} 
This work has been supported through PRIN MIUR 2007 
``Multiple stellar populations in globular clusters: census, characterization and
origin". We thank S. Cassisi for useful comments.
We also warmly thank the anonymous referee for his careful
reading of the first version of the manuscript, and for forcing us to clarify a number of
important points.

\label{lastpage}

\end{document}